\documentclass[journal]{IEEEtran}

\usepackage{xcolor}
\usepackage{grfext}
\usepackage{kvoptions}
\usepackage{kvsetkeys}
\usepackage{adjustbox}
\usepackage{collectbox}
\usepackage{ltxcmds}
\usepackage{cite}
\usepackage{subfigure}
\usepackage{amsmath,amssymb}
\usepackage{textcomp}
\usepackage[normalem]{ulem}
\usepackage{epstopdf}
\usepackage{adjustbox}
\usepackage{mleftright}
\usepackage{flushend}

\definecolor{mycolor}{RGB}{0, 0, 0}

\begin{document}

	\title{Passive Microwave Tag Classification Using RF Fingerprinting and Machine Learning}

	\author{Cory Hilton,~\IEEEmembership{Graduate Student Member,~IEEE},
 Mohammad Rashid,~\IEEEmembership{Member,~IEEE},
 Faiz Sherman, Steven Bush,
 and Jeffrey A. Nanzer,~\IEEEmembership{Senior Member,~IEEE}
		\thanks{Manuscript received 2025.}
		\thanks{This work was supported in part by the Procter \& Gamble Company. \textit{(Corresponding author: Jeffrey A. Nanzer)}}
		\thanks{C. Hilton and J. A. Nanzer are with the Department of Electrical and Computer Engineering, Michigan State University, East Lansing, MI 48824 USA (email: hiltonc2@msu.edu, nanzer@msu.edu).}
		\thanks{M. Rashid is with the Department of Electrical and Computer Engineering \& Technology, Minnesota State University, Makato.}
		\thanks{F. Sherman and S. Bush are with the Procter \& Gamble Company.}
	}

	\maketitle
	
	\begin{abstract} 

We present an approach to identifying wireless microwave tags using radio frequency (RF) fingerprinting and machine learning. The tags are designed for low cost and simplicity, consisting of only two antennas and a single nonlinear element (a diode). An interrogating transceiver transmits a signal consisting of a set of individual frequency tones that is captured by the tag. The signal response of the diode is nonlinear, and can be represented by an infinite power series, the coefficients of which are similar but not identical for different physical diodes due to small manufacturing perturbations. The small differences in the signal responses manifest in the spectral signal response of the tag, which is retransmitted back to the interrogating transceiver. Input into machine learning algorithms, the slight differences in the spectral responses of the diodes can be used to uniquely identify devices. To demonstrate the concept, we designed 2.0 GHz tags consisting of patch antennas and a single diode, along with a bi-static radar system operating at the 2.0 GHz 802.11 Wi-Fi band transmitting multi-tone continuous wave signals representing common 802.11 training fields. The received signals were processed using a set of algorithms for comparison purposes. A real-time classification accuracy of $95\%$ between two tags was achieved.

	\end{abstract}

	\IEEEpeerreviewmaketitle

	\section{Introduction}

	

Rapid expansion of the Internet of Things (IoT) has led to a significant increase in wirelessly connected and monitored devices. As technologies become increasingly connected, {physical layer security has become a critical aspect} of ensuring reliable functionality \cite{4801689,7123188,10286341}. In particular, accurate identification of devices, whether or not they are equipped with smart functionality, is crucial for applications like human-computer interfacing, home health, and consumer-product interactions, among others \cite{9947030,7836336}. Radio-Frequency Identification (RFID) is an established approach to identifying and tracking objects and devices by using radio frequency (RF) tags attached to objects along with an interrogating transceiver that emits a signals and monitors the signals retransmitted by the tag \cite{8970312,8041691}. Typically, RFID tags include designed signatures which may be encoded using on-tag circuitry \cite{9050631}; these signatures allow the unique identification of devices for tracking purposes. However, the methods of encoding unique signatures generally entails the use of complex circuitry to impart the signatures. These may be clever uses of passive elements or active circuitry, which subsequently requires on-board power or energy harvesting \cite{9049125,807936}; in either case the cost increases, making scalability and utility on "dumb" devices lacking smart device connectivity increasingly infeasible. 

RF fingerprinting is a technique that uses the inherent, minute signatures of electronics to differentiate between devices \cite{10592579,9913208,10318021}. The voltage response of nonlinear components, such as a diode, can be represented by an infinite power series, the coefficients of which are dependent on the specific device, for example whether it is active or passive \cite{9913208,812242,10795430}. Given a set of manufactured devices of the same type, the magnitude of these coefficients is largely the same. However, small imperfections in the materials and manufacturing process yield small differences in the signal responses of each device, even when they are of the same design. These differences manifest in the power series representation and minute perturbations on the coefficients. By appropriately training a classification algorithm, these minute differences can be used to uniquely identify the device based on its signal response.

RF fingerprinting has long been studied as a method of device identification and localization for a wide range of applications \cite{8970312,10008216}. 
The concept is based on the unique signatures present in the signals emitted by nonlinear electronic devices in integrated circuits, which are specific to the non-idealities of the specific chip, enable unique identification capabilities~\cite{8041691}. 
RF fingerprinting techniques has been widely developed and used in applications including health sensing, precision construction and agriculture, and device identification~\cite{10654361,9103056,10130539}; however, chip-based fingerprinting can have limitations that limit the SNR capabilities of passive approaches, and thus recently chipless approaches have been explored through time and frequency domain analysis of the responses of designed RF tags~\cite{9050631,9697091}. Such methods have shown that localization, range, and device identification can be performed in conjunction with classification~\cite{8970312,10722864,10136207}
While the above recent work has shown significant promise for RF fingerprinting uses, applicability to non-smart devices where active component responses may not be available has not been assessed. Furthermore, the co-design of tags and the monitoring system is an approach affording significant design opportunities that has not been fully explored.

\begin{figure*}[t!]
\begin{center}
    \centering
   \includegraphics[width=0.75\textwidth]{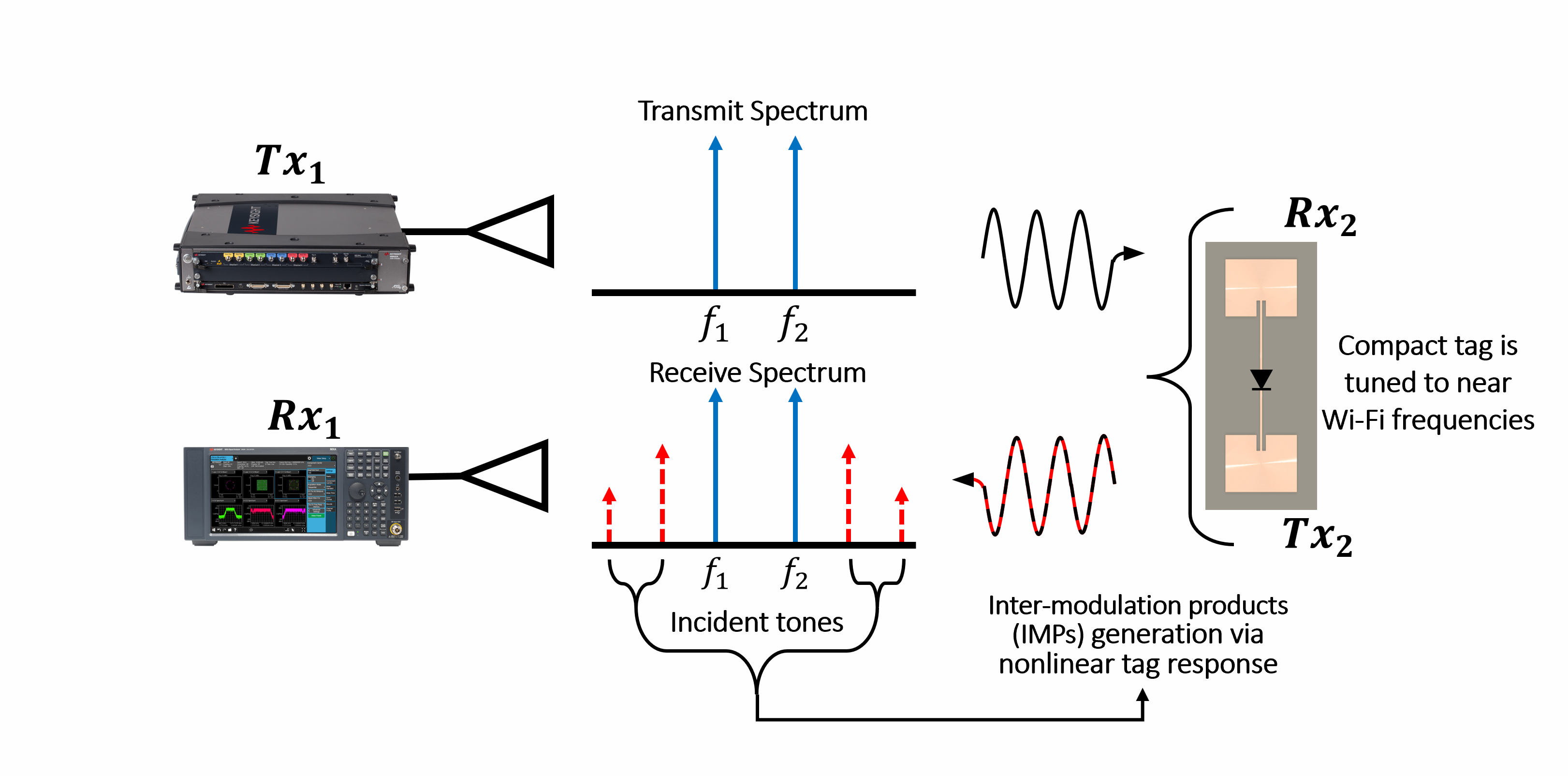}
  \caption{Usually, the classification of passive tags equipped with nonlinear COTS devices is enabled via the extraction of unique spectral information provided by each device. Tags are tuned near Wi-Fi frequencies to allow for existing 802.11 training fields to be used as a method of exploiting existing signals within the environment. Each tag receives the signal(s) at the fundamental frequency, which are then modulated by the nonlinear device, producing harmonics. The modulated signal is then re-transmitted with a response with model-varying harmonic power levels that can be detected independent of clutter within an environment. This is due to the minute manufacturing defects that exist within each low-cost device, yielding spectra with differing power levels as well as inter-modulation product order ratios. These spectra can then be processed for change detection and device presence classification.}
  
\label{Fig - OV1}
\end{center}
\end{figure*}

In this paper, we demonstrate the use of RF fingerprinting on low-cost, simple RF tags consisting of only passive antennas and a single diode. Our approach is based on illuminating the tags with signals consisting of a set of frequency tones such as those present in common Wi-Fi training fields. The signal containing the tones, which are transmitted simultaneously, is received by the tag and passed through the diode. The nonlinear response of the diode thus creates intermodulation products due to the presence of multiple frequencies in the signal. The peak power of the intermodulation products are processed by the receiver as a function of time and used to train machine-learning (ML) classification algorithms. We fabricated two tags, each consisting of two patch antennas operating at 2.0 GHz connected in between by a single Schottky diode. We created a bistatic transceiver consisting of a transmitter emitting the multi-tone signal and a signal analyzer capturing and demodulating the received signals. A real-time classification accuracy of $95\%$ between two tags was achieved.

\section{Nonlinear Device Signatures}                      \label{S-theory}

\subsection{Nonlinearity-Based RF Fingerprints}

RF fingerprints are features in RF signals that can be used to uniquely identify the physical device that transmits the signal. Successful identification is based on adherence to the following properties~\cite{10.1145/2379776.2379782}
\begin{itemize}
\item 
\textit{Universality:} Every device should have a fingerprint.
\item
\textit{Uniqueness:} The fingerprint should be different for each device.
\item
\textit{Permanence:} The fingerprint should be invariant over time.
\item
\textit{Collectability:} A receiving system should be able to detect the fingerprint.
\end{itemize}
The set of potential RF fingerprints generated by wireless systems is expansive, however one of the most directly obtained, and one that best matches the above properties, is the response of nonlinear components in the hardware. Because of small differences in the devices due to finite manufacturing tolerances, the nonlinear signature of each component will be slightly different, which manifests in signals passed through the device. 

The voltage response of a nonlinear device to an input voltage $v_i$ can be represented by a Taylor series\cite{cripps2006rf}
\begin{equation}
v_0 = a_0 + a_1 v_i + a_2 v_i^2 + a_3 v_i^3 + \cdots = \sum_{n=0}^{\infty}{a_n v_i^n}
\end{equation}
where the coefficients are given by
\begin{equation}
a_n = \mleft\vert\frac{d^n v_o}{dv_i^n}\mright\vert_{v_i = 0}
\end{equation}
The coefficient $a_0$ represents the dc voltage output, while $a_1$ is the coefficient of the linear output of the signal. The additional terms manifest the nonlinear response of the device. For a monochromatic inputs signal $v_i = A\cos\left(\omega_0 t\right)$ where $A$ is the signal amplitude and $\omega_0 = 2 \pi f_0$ is the angular frequency, the output voltage of the nonlinear device for terms up to third order is 
\begin{equation}
v_o = a_0+ + a_1 A \cos\left(\omega_0 t\right) + a_2 A^2 \cos^2\left(\omega_0 t\right) a_3 A^3 \cos^2\left(\omega_0 t\right)
\end{equation}
which expands to
\begin{equation}
v_o = \left(a_0 + \frac{1}{2}a_2 A^2\right) + \left(a_1 A + \frac{3}{4}a_3 A^3\right)\cos\left(\omega_0 t\right) + \cdots
\end{equation}
The amplitude of the output signal at the fundamental frequency $\omega_0$ is thus 
\begin{equation}\label{fund_amp}
a_1 A + \frac{3}{4}a_3 A^3
\end{equation}
and the gain of the signal at the fundamental is \eqref{fund_amp} divided by the input signal $A\cos\left(\omega_0 t\right)$,
\begin{equation}
a_1 + \frac{3}{4}a_3 A^2
\end{equation}

The amplitude of the signal passed through a given nonlinear device up to third order is thus modified by the two coefficients $a_1$ and $a_3$. Subsequent odd-order coefficients also impact the signal, albeit with decreasing impact. Because of manufacturing tolerances, the coefficients $a_n$ for a specific device, such as a diode, will be the similar to first order but increasingly different at higher orders, such that coefficients may vary by a few percent. In practice, this has little effect on the overall operation of the devices, since the main contributing component $a_1$ is essentially the same. However, even small variations in the coefficients due to manufacturing tolerances manifest small changes in the signal response that can be detected and classified using machine learning algorithms.

\begin{figure}[t!]
    \centering
    \includegraphics[width=\linewidth]{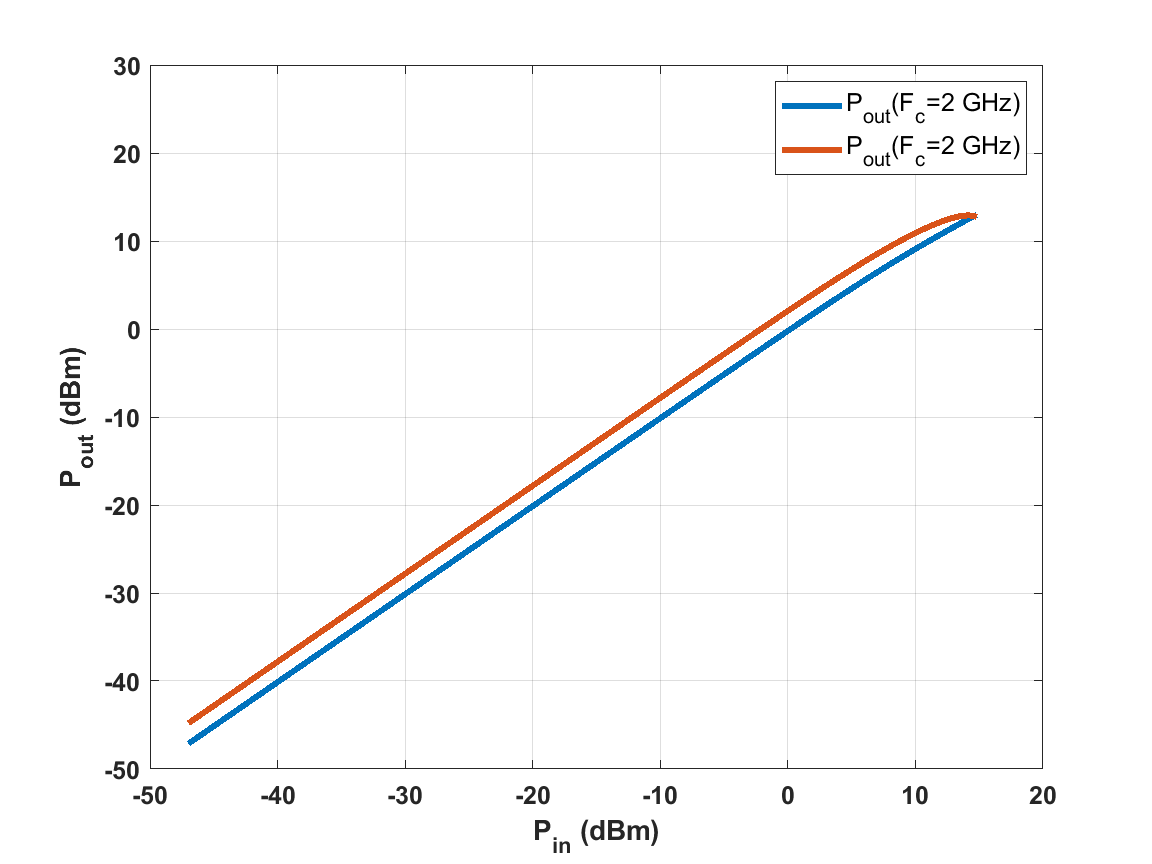}
		
		(a)
		
		\includegraphics[width=\linewidth]{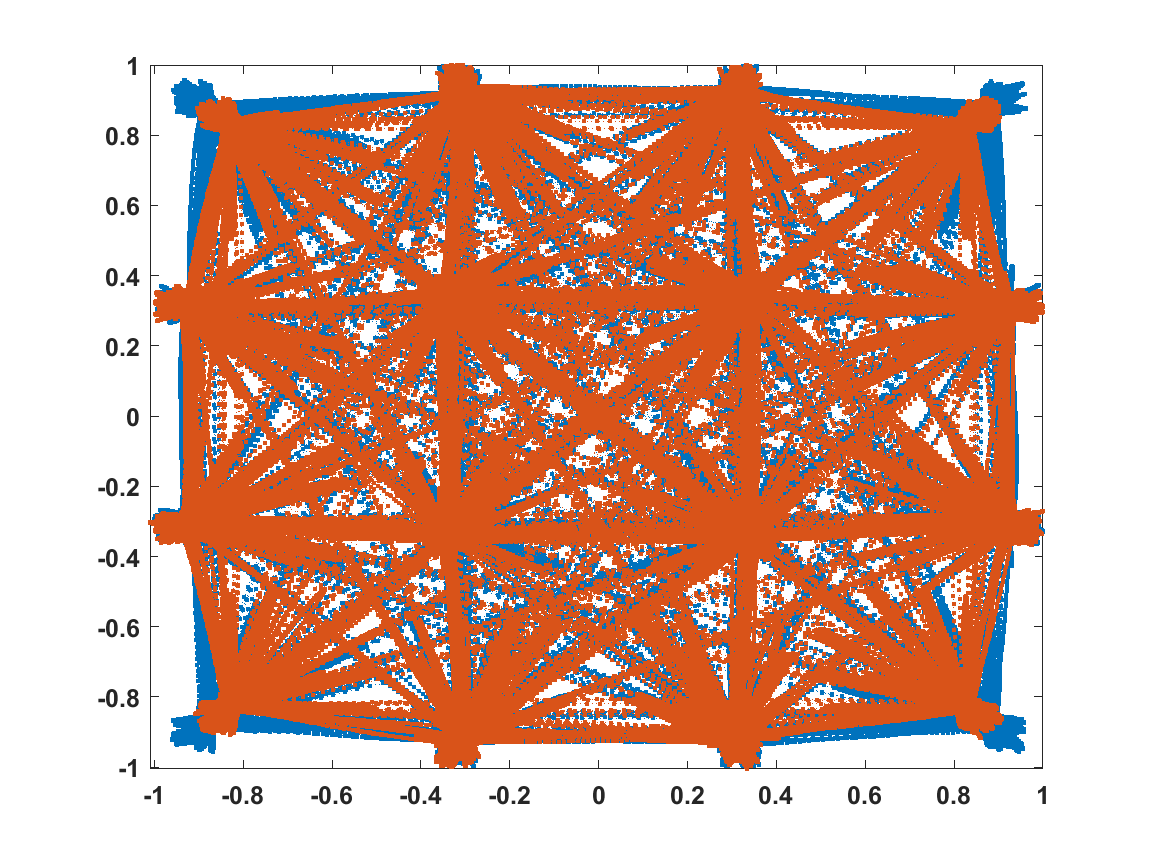}
		
		(b) 
		
    \caption{(a) Output power of a theoretical amplifier with random $\pm$ 5$\%$ modulation of the high order Taylor series coefficients. The results of which demonstrate that even small variance in the harmonic amplitudes of a non-linear device can result in large ($ 3 $dB) differences in device behavior. (b) Impacts of the varying non-linear coefficients on a common 16-QAM signal. The symbols at each location noticeably shift due to the presence of these harmonic impacts, thus enabling potential methods of classification to be introduced for device identification.}
    \label{amp_curves}
\end{figure}

As an example, we consider a general model of an RF amplifier, because the impact of differences in the model can be readily discerned. We represent an arbitrary amplifier using the Taylor series
\begin{equation}
v_0 = v_i - 0.333v_i^3 + 0.133 v_i^5 - 0.05v_i^7 + 0.022v_i^9
\end{equation}
The coefficients were varied randomly by 5\% to generate the two responses shown in Fig.~\ref{amp_curves}. The input-output power curves are shown in Fig.~\ref{amp_curves}(a), which show small differences in the overall shape. While the gain differs between the two models, this difference results mainly in a higher or lower power received by the system, and thus is not a persistent property which could be considered a fingerprint. However, the impact of the higher order nonlinearities can be seen by considering a 16-QAM (quadrature amplitude modulated) signal passed through the models, since the compression point of the amplifiers is slightly different, as seen by where the output power of the two models begins to converge in Fig.~\ref{amp_curves}(a). The constellation diagram of the signal for the two models is shown in Figs.~\ref{amp_curves}(b) and (c). The normalized constellations show significant differences near the corners of the diagram due to the different nonlinear responses; this is anticipated, since the complex signals has its highest amplitude (highest peak-to-average ratio) at the corners of the diagram. While these differences are small and would not significantly impact demodulation of the data, the differences are persistent since they are due to the physics of the device and can thus be used as an RF fingerprint. 

\subsection{Nonlinear Tag Design}

The above model is adequate for modeling the effects of a weakly nonlinear device, however it is a limited model; for example, no explicit phase information is represented in the signal output. However, it serves as an example of the impact of small differences in the nonlinear response of devices, however in this work we seek to identify differences due to the nonlinear response of diodes, which manifests in less obvious ways than the compression of an amplifier. We consider the input signal to be a superposition of multiple input tones, given by
\begin{equation}
v_i = A\cos\left(\omega_1 t\right) + B\cos\left(\omega_2 t\right)
\end{equation}
After passing through the nonlinear response, the resultant signal includes components at a series the frequencies
\begin{equation}
m\omega_1 \pm n\omega_2,\; m = 0,1,2,\dots,\; n = 0,1,2,\dots
\end{equation}
We are interested in this work in the signals manifesting at the fundamental frequencies $\omega_1$ and $\omega_2$, and in the nearest intermodulation products.

\begin{figure}[t!]
\centering
{\includegraphics[width=\linewidth]{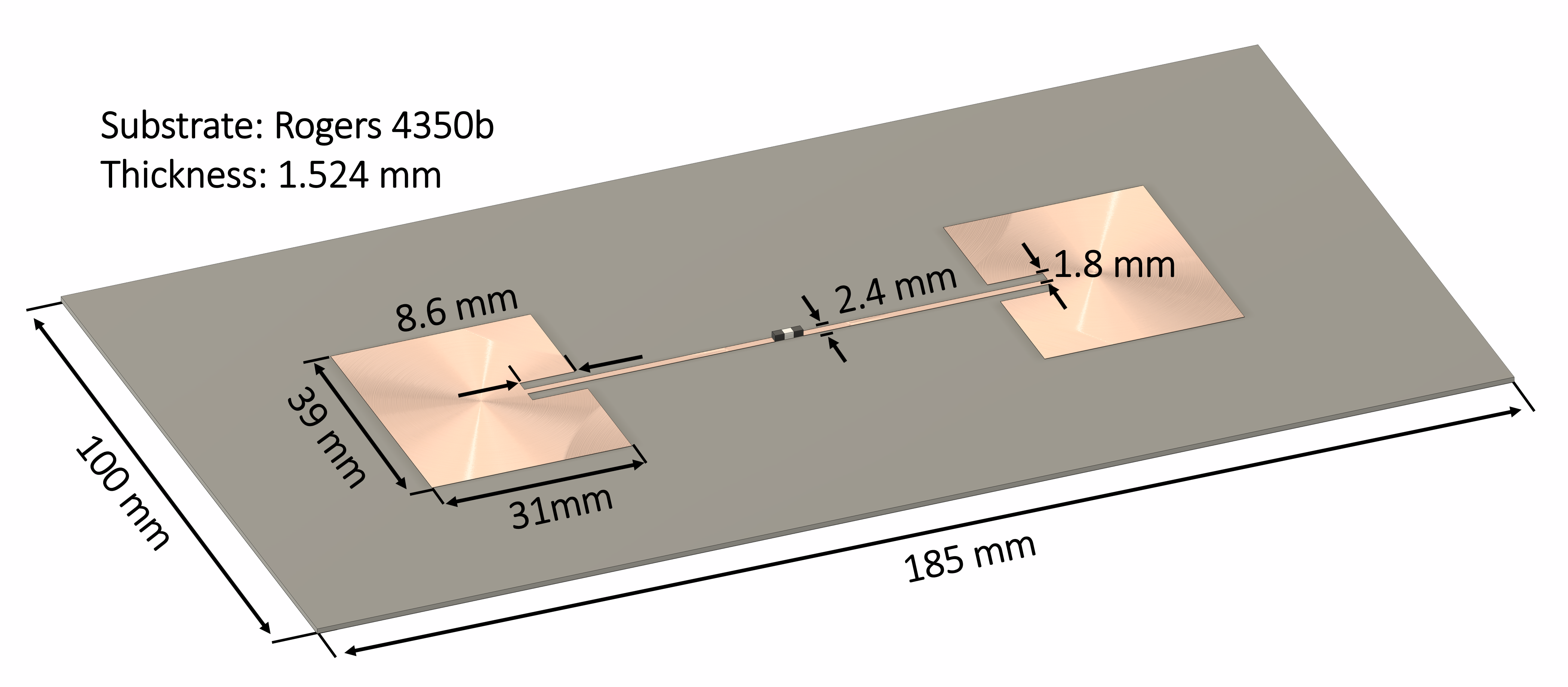}}

\caption{Model of the nonlinear tag used for this work. Designed on Rogers~4350b with a thickness of 1.524 mm, two patch antennas were designed at a center frequency of 2.0 GHz with a BAT15-03W Shottky diode attached to the ports between the antennas.}
\label{Antenna}
\end{figure}

We explore the concept using a simple passive tag consisting of two patch antennas connected by a diode. The dimensions of the antennas were 3.8 cm in width and 2.5 cm in height which yielded a resonance frequency of 2.0 GHz and a maximum broadside realized gain of 5.81 dBi. A BAT15-03W diode was placed as a connector between each of the two antennas, acting as a shared port. Two tags were fabricated on 1.542 mm Rogers 4350b substrate with a microstrip feedline width of 3.03 mm. The tag design can be seen in Fig.~\ref{Antenna}.

\begin{figure}[t!]
\centering
{\includegraphics[width=\linewidth]{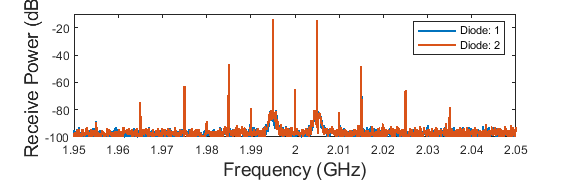}} 

(a)

{\includegraphics[width=1\linewidth]{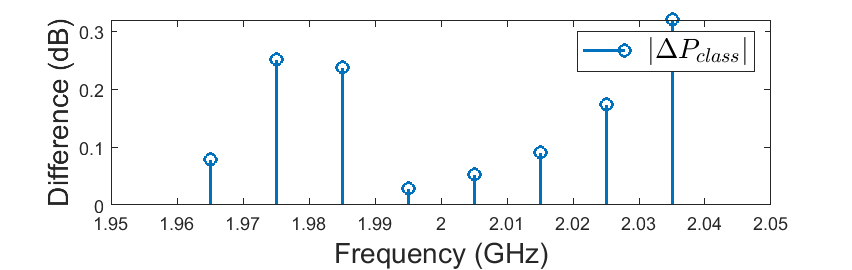}} 

(b)

%

\caption{(a) Example spectra for nonlinear tag 1 and nonlinear tag 2. In both cases the tags were manufactured identically using the general patch antenna designed at 2.0 GHz. The spectra have minimal apparent differences. (b) The difference between the two spectra over an average of 1000 FFT samples; 3rd order and higher IMPs exhibit significant variation. 
}
\label{spectra_plot}
\end{figure}

The nonlinear response of the tags was measured wirelessly using a bistatic setup with a Keysight E8267D vector signal generator transmitting a two-tone signal at frequencies $f_1 = 1.95$~GHz and $f_2 = 2.05$~GHz through an 8 dBi log-periodic antenna. The signal was transmitted to the tag at a distance of 50 cm, and the signal retransmitted by the tag was captured using a Keysight MSO-X-9200A oscilloscope with an 8 dBi log-periodic antenna.
We show examples of measurements of the two fundamental frequencies and the two nearest intermodulation products, $\omega_1$, $\omega_2$, $2\omega_1 - \omega_2$, and $2\omega_2 - \omega_1$, in Fig.~\ref{spectra_plot}. The spectra are superficially similar, however the minor differences in the peaks of the signals (indicated with red dots) will be slightly different because of the different nonlinear response of the diodes. These minor differences are the basis for the classifier discussed in the next section.

\section{Experimental Setup}                     \label{S-perturb}



\begin{figure}[t!]
    \centering
    \includegraphics[width=0.6\linewidth]{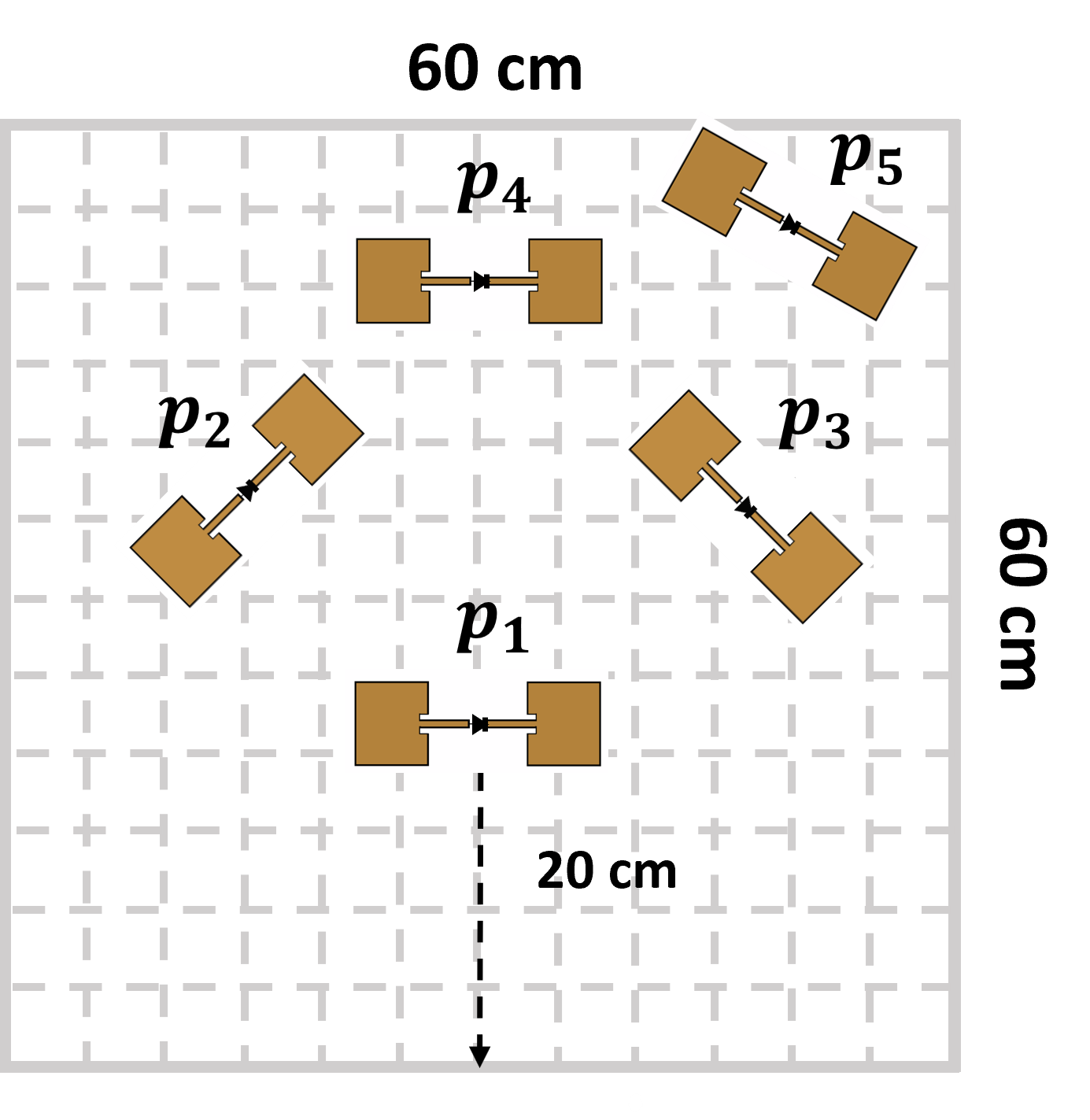}
    \caption{Grid displaying the positions of the large and small perturbation measurements taken for the passive nonlinear tag. The small perturbation measurements considered only displacement and replacement of the tag at position $P_{1}$. The large perturbation measurements consider all locations for a complete analysis of system tolerance within a range of 60 cm, using all positions $P_{1}$ through $P_{5}$ for testing.}
    \label{fig:model -perturbation grid}
\end{figure}

To evaluate the ability to detect and classify separate tags based on their RF fingerprint, experiments were conducted using the setup described in the last section to collect signals from the tags. The tags were moved to various locations within a 60$\times$60 cm space to evaluate the performance under varying multipath conditions, varying SNR, and varying angles of the tag. The positions are shown in Fig.~\ref{fig:model -perturbation grid}, and the positional values are given in Table~\ref{tab:my_label}. We collected three different data sets with which to evaluate a set of classifiers, described as follows.

\begin{table}[t!]
\centering
\caption{Positions of the passive nonlinear tag within the 60 cm by 60 cm grid during the large perturbation measurement process.}
\label{tab:my_label}
\resizebox{\linewidth}{!}{ 
\begin{tabular*}{\linewidth}{||c|ccccc||}
\hline
\textbf{Position} & \textbf{1} & \textbf{2} & \textbf{3} & \textbf{4} & \textbf{5} \\ \hline
Grid position (cm) & (30,20) & (43,45) & (17,45) & (30,55) & (53,53) \\ \hline
\end{tabular*}
}
\end{table}

\vspace{12pt}
\begin{enumerate}
    \item \textit{Static Case:} The static case featured the nonlinear tag being placed in a single position and remaining unmoved for the duration of the training and testing processes. In subsequent cases, the measurements of the static case were considered to be the control measurements. 1001 samples/position were collected. 
    \item \textit{Small Perturbation Case:} The small perturbations can be described as a series of micro-motions of the tag. The tag was picked up and replaced in the same position at location $P_{1}$ repeatedly, creating minor positional variations. 1001 samples/position were collected for each of five lightly perturbed locations.
    \item \textit{Large Perturbation Case:} In the large perturbation case the tag was picked up and replaced at designated positions and angles relative to the transmit and receive antennas of the system as shown in Fig.~\ref{fig:model -perturbation grid}. 1001 samples of each diode/position were measured prior to training of each classifier model and testing.
\end{enumerate}

\section{Machine Learning Based Classification of Diodes Using 
Frequency Spectra}   \label{S-dsp}

In this section, we demonstrate the classification performances of 
several different ML algortihms for 
classifying diodes using their wireless frequency spectra. 
In order to reduce the dimensionality of the feature space as well as 
the complexity of the ML algorithms, 
only a small number of significant 
features were selected from the spectra. For that purpose, the spectra 
were processed through the peak detection algorithm \cite{peak_detect} which 
outputs the values of the four highest peaks in each spectrum. 
The peak values based dataset contained samples for the static case, 
the small perturbation case, and the large perturbation case. 
Particularly, the static case (no perturbation) 
dataset contained $1007$ samples for each of the 
two classes and it was split into the training and test sets using a $90-10$ split. 
The training set for the static case dataset 
was used to train the ML classifiers unless stated otherwise, whereas the learned 
classifiers were then tested on the unseen data for all the 
considered perturbations of diodes including the static case. 
Prior to classification, 
the entire dataset was also standardized \cite{scikit-learn} by fitting on the static case training data whereas the learned parameters were used 
for transforming the small perturbation and the large perturbation datasets. 
Moreover, 
the labels were assigned to the training examples using the K-means 
algorithm \cite{scikit-learn} with $K=2$, and 
then the synthetic minority over-sampling technique \cite{SMOTE_2002} 
was used to balance the labeled dataset. 

We consider 
classifiers such as perceptron and logistic regression which have been shown to be highly capable of classifying linearly separable classes, making them useful in cases like change detection algorithms. Additionally, various other classifiers including the support vector machine (SVM), decision tree, random forest, K-nearest neighbor (KNN), and Gaussian process classifier (GPC) which can classify the non-linearly separable classes as well are also explored. In this case, the KNN and GPC classifiers were chosen based on their weighted distribution models, allowing for analysis on whether change detection could be performed with prior knowledge of existing samples within the scene. The decision tree and random forest classifiers were chosen for the ability to have a highly modifiable set of models allowing for the number of weights and features in each sample to be well described and weighted accordingly, allowing for the results to demonstrate the effectiveness of change detection when presented with as much information as is present in each feature set. Finally, the SVM was considered because of its ability to classify highly linearly separable datasets. This extends into the the one-class support vector machine (OcSVM) as well, in which the output of the model is further reduced to either a true or false return, making this an idea model for simple change detection that, while accurate, may increase the volitity when regarding sample consistency. The scikit-learn library to implement these classifier models in Python \cite{scikit-learn}.

We evaluate the classification performances of the above 
classifiers for the following cases:
\begin{enumerate} 
  \item The models are trained on the 
static case training set and generalized to the test sets of static case, small 
perturbation, and large perturbation, 
  \item The models are trained on an increased-sample training dataset and 
generalized to the static, small perturbation, and large perturbation 
test sets.
  \item The models are trained on only the static increased-sample training dataset and generalized as a function of both small and large perturbation cases. 
\end{enumerate}

\begin{figure*}[h!]
    \centering
    \begin{tabular}{ccc}
    \subfigure[]{\includegraphics[width=0.33\linewidth]{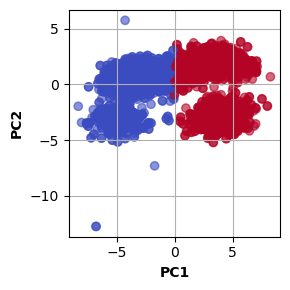}} 
  
    \subfigure[]{\includegraphics[width=0.33\linewidth]{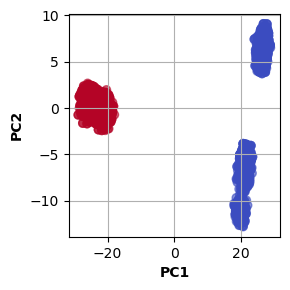}} 
    
    \subfigure[]{\includegraphics[width=0.33\linewidth]{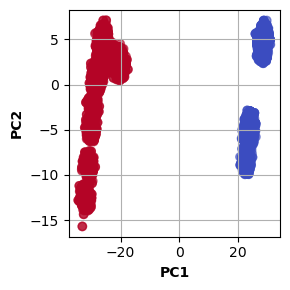}} 
     
    \end{tabular}

\caption{Principal components of the three considered data set (a) PCA output training on data collected from static tags, (b) PCA output training on a combined dataset of static and tags that were moved only over short ranges to demonstrate robusticity of the classifier to the antenna patterns of the tag, and (c) PCA output training on a combined dataset of static and perturbed tags of both the small ($<$1 cm) and large ($>$10 cm) ranges. While overlap between the clustered responses of class 0 and class 1 are present, there remains a clear linear separation between the classes that allows for low-computational complexity classifiers to accurately predict the tag response based on a set of peak values from the observed spectra.}
		\label{PCA}
    \end{figure*}

\begin{figure*}[t!]
    \centering
    \begin{tabular}{ccc}
    \subfigure[]{\includegraphics[width=0.16\linewidth]{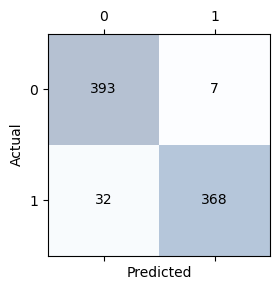}} 
  
    \subfigure[]{\includegraphics[width=0.16\linewidth]{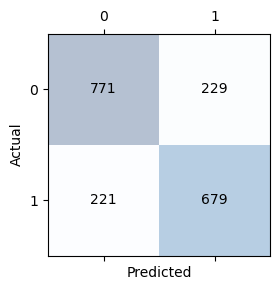}} 
    
    \subfigure[]{\includegraphics[width=0.16\linewidth]{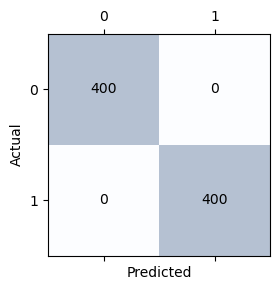}} 
   
    \subfigure[]{\includegraphics[width=0.16\linewidth]{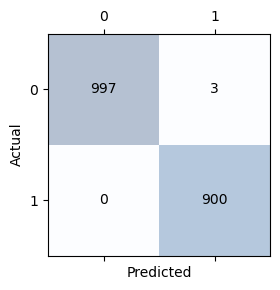}} 
    
    \subfigure[]{\includegraphics[width=0.16\linewidth]{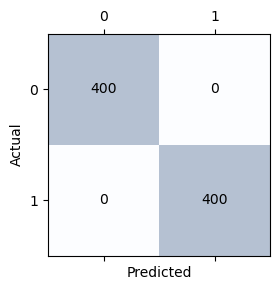}} 
    
    \subfigure[]{\includegraphics[width=0.15\linewidth]{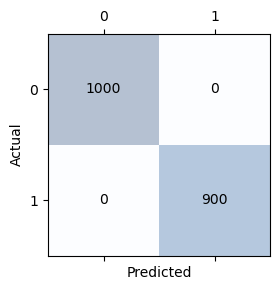}} 
     
    \end{tabular}
    
    \caption{Confusion matrices for the test cases of the three classifiers displaying accuracy's greater than $90\%$ for both the large and small perturbation cases of the nonlinear tag with the exception of the perceptron for the case of larger perturbations. The left column displays those matrices belonging to the small perturbation case and the right belonging to those model used in the large perturbation case. Figures (a) and (b) show the accuracy of the perception model, (c) and (d) show the accuracy of the decision tree model, (e) and (f) show the accuracy of the random forest model.}
		\label{confusion_matrices}
    \end{figure*}

    
		\begin{table*}[t!]
        \centering
        \caption{\textbf{ Training classifiers on the static case only}: 
				Classification accuracies for the static case (no perturbation), the small 
				perturbation, and the large perturbation cases.}
        \label{tab:static_training}
        \begin{tabular}{||c||c|c|c|c|c|c|c||}\hline
           \textbf{Case} & \textbf{Perceptron} & \textbf{Logistic Regression }& 
					\textbf{SVM} & \textbf{KNN} & \textbf{GPC} & \textbf{Decision Tree} &
					\textbf{Random Forest}\\ \hline
					 Static Case        & 99.9\% & 99.9\% & 99.9\% & 99.9\% & 99.9\% & 99.9\% & 99.9		
					\%\\ \hline
           
           Small perturbation & 99.9\% & 99.8\% & 95.0\% & 98.3 \% & 89.5 \% &  99.9\% & 				99.9\% \\ \hline					
           Large perturbation & 93.4\% & 97.1\% & 97.4\% & 93.8\% & 87.8\%& 99.0\%& 99.8\%
					\\ \hline 					
        \end{tabular}
    \end{table*}

    \begin{table*}[t!]
        \centering
				        \caption{\textbf{ Training classifiers on an increased-sample dataset}: 
								Classification accuracies for the static case (no perturbation), the 
								small perturbation, and the large perturbation cases.}
        \label{tab:increased-sample_training}

        \begin{tabular}{||c||c|c|c|c|c|c|c||}\hline
           \textbf{Case} & \textbf{Perceptron} &\textbf{ Logistic Regression} & 
					\textbf{SVM} & \textbf{KNN} & \textbf{GPC} & \textbf{Decision Tree} &
					\textbf{Random Forest}\\ \hline
					 Static Case        & 99.9\% & 99.9\% & 99.9\% & 99.9\% & 99.9\% & 99.9\% & 99.9		
					\%\\ \hline
          
           Small perturbation & 97.6\% & 98.4\% & 99.3\% & 99.8\% & 99.7\% &  99.8\% & 99.9\% \\ \hline					
           
           Large perturbation & 76.3\% & 80.1\% & 85.7\% & 98.9\% & 88.6\%& 99.8\%& 
					99.9\%
					\\ \hline 				
        \end{tabular}
    \end{table*}


\subsection{Training Classifiers on the Static Case Only}
In Table \ref{tab:static_training}, we present the classification 
performances of the above classifiers when 
trained on the static case training set 
and generalized to the static case test set, 
the small perturbation 
test set, and the larger perturbation test set. 
For the static case, the test set contains $100$ samples per class, 
whereas both the small perturbation and the large 
perturbation test sets contain $220$ samples per class. 
It is observed that 
as the classes are highly linearly separable 
in the static case, all the 
classifiers demonstrate nearly $100\%$ classification accuracy. 
However, in the cases of small perturbations and large perturbations 
we observe a slight degradation in classification performance. 
Particularly, while almost all of the classifiers can easily generalize to 
the small perturbation dataset with about $99\%$ accuracy, however, 
some of them show degraded performances in the large perturbation case 
due to convergence errors in 
their optimization processes. Nonetheless, the simple perceptron 
classifier and tree-based classifiers outperform others in 
all cases with about $94\%$ classification accuracy. This 
is also observed for the perturbed dataset, 
wherein we observe that these classifiers 
demonstrate reduced misclassification errors compared to others particularly 
in the large perturbation case.

\subsection[(b)]{Training Classifiers on an increased-sample Dataset}
Here, we consider two different training sets for the small perturbation and 
large perturbation cases each with $1814$ samples ($907$ samples per class). 
We concatenate both datasets with the static case training set to 
train the classifiers 
on the increased-sample dataset. Table \ref{tab:increased-sample_training} demonstrates 
the classification performances of the classifiers when generalized to the 
same test sets as were used in Table \ref{tab:static_training}. It is observed 
here that, in the static case, 
the addition of perturbed data in the training set degrades 
the classification performances of the perceptron and the 
logistic regression classifiers, whereas in the small perturbation case, 
all the classifiers except SVM show poor 
performances.The increased performance 
of SVM can be attributed to its ability of learning the decision region 
that has the largest separation between the closest training points of any 
class. This is well described by the one-dimensional linear model of the SVM,
\begin{equation}
E_{SVM} = w^{T}x_{i}-b
\end{equation}
where this is expanded for the nonlinear model as
\begin{equation}
E_{SVM} = \frac{1}{n}\sum_{i=1}^{n}{\zeta_{i}+\lambda||w||^{2}}
\end{equation}
where
\begin{equation}
\zeta_{i} = \mathrm{max}(0,1-w^{T}x_{i}-b)
\end{equation}
and $w^{T}$ is the transposed weighting matrix, $x_{i}$ is the observed sample, $b= -0.77$ is a constant offset, and $\lambda$ is selected to force hard-margin separability from a search over the range $\lambda = 10^{[-3:2]}$ to achieve the highest accuracy for the models shown.

In the case of large perturbation we observe in 
Table \ref{tab:increased-sample_training}
that the performances of all the classifiers improve significantly 
and that some of them show $99.8\%$ accuracy as compared to the performances 
in Table \ref{tab:static_training} where only the static case training set 
was used for training of classifiers. 
However, the SVM classifier herein 
Table \ref{tab:increased-sample_training} shows the best performance 
for all the considered perturbation and 
static cases but achieves only $94.1\%$ accuracy in the 
large perturbation case.
This accuracy of SVM was matched 
closely in Table \ref{tab:static_training} by the perceptron and the tree-based 
classifiers where the static cases training set was used to train the classifiers.

\subsection{Training Classifiers on Combined Large Datasets}
A similar trend can be seen in Table V, where the number of samples in relation to Table IV has increased to 1001 samples/position per class. Yielding a total of 5,005 samples, this model is now trained on the entirety of the static case due to assumed quasi-static environments in practical application of such classification methods. The results can be shown to have reduced classification accuracy when compared to the decimated training datasets. This is likely due to the varying amplitudes introduced by the perturbations cases varying the amplitudes of the resultant nonlinear responses, thus yielding false responses from each classifier. The results of a principal component analysis of the three data sets is shown in Fig.~\ref{PCA}, where it can be seen that the static data set shows a clearer separability than the dynamic or combined data sets. However, the perceptron and tree-based classifiers are shown to have high accuracy greater than $90\%$ due to their weighted dependencies on the samples near that of the observed sample as well as a reduced number of observed features in each sample. This is further corroborated in the confusion matrices of these results as shown in Fig.~\ref{confusion_matrices}.

To summarize, the above analyses show that in general training the classifiers 
only on the static case data generalizes well to the perturbed datasets. Furthermore, 
it represents the most common scenario 
seen for a longer time by the base station,  
as compared to the small perturbation and the large perturbation scenarios 
which are also harder to detect. In other words, the 
static case can be observed by the base station 
for a long time duration which can be leveraged to well-train the ML classifiers. 
As such, in the following, only training classifiers on the static case training data is assumed.

\section{Conclusion}

In this paper we demonstrated the use of machine learning algorithms to classify RF fingerprints generated by minor manufacturing differences in nonlinear electronic components. Through a simple analysis of the peaks in the retransmitted spectrum of an RF tag with a diode, the minor differences in the nonlinear transfer function of the diode can be detected and used for classification. {The experiments in this work focused on diode signatures. However, the approach is broadly applicable to the classification of nonlinear devices such as amplifiers or other components.} Furthermore, the simplicity of the RF tag in this work highlights the ability to perform RF fingerprinting in a low-cost manner, using simple signals consisting of only a few discrete frequency tones. 
	

	\bibliographystyle{IEEEtran}
	\bibliography{nb}

\end{document}